# Importance of reorientational dynamics for the charge transport in ionic liquids

P. Sippel, S. Krohns, D. Reuter, P. Lunkenheimer[*], and A. Loidl

*Experimental Physics V, Center for Electronic Correlations and Magnetism, University of Augsburg, 86159 Augsburg, Germany*

Most ionic liquids contain at least one rather complex ion species exhibiting a dipolar moment. In the present work, we provide a thorough evaluation of broadband dielectric spectra of 12 ionic liquids taking into account the often neglected reorientational dynamics of these ions. We confirm that this dynamics leads to a clear relaxational signature in the spectra, a fact that so far only was considered in few previous works. The obtained reorientational relaxation times are well consistent with earlier inelastic light-scattering and high-frequency dielectric investigations. Evaluating our dielectric spectra in terms of reorientational motions reveals a close coupling of the ion-rotation dynamics to the ionic charge transport in a broad temperature range from the low-viscosity liquid above room temperature deep into the high-viscosity supercooled state close to $T_g$. This coupling does not seem to be mediated by the viscosity but probably is of more direct nature, pointing to a revolving-door mechanism as also considered for plastic-crystalline ionic conductors. Our results show that the reorientational motion of the dipolar ions plays a significant and so far widely overlooked role for the ionic charge transport in ionic liquids.

## I. INTRODUCTION

Molecular reorientation dynamics plays an important role in many material classes. This includes, e.g., molecular glass-forming liquids, where the glass transition was suggested to be governed by the freezing of orientational degrees of freedom, playing the dominant role compared to flow processes [1,2]. Further prominent examples are the so-called plastic crystals, where the molecules can reorient, while their centers of mass are fixed on a crystalline lattice [3]. These materials exhibit a glass transition where the orientational degrees of freedom slow down, while translational motion plays no role at all. Recently, these molecular reorientations have gained considerable interest because they seem to strongly favor the exceptionally high ionic conductivity of some plastic crystals [4,5] via a "revolving-door" like mechanism [6,7]. Ionic liquids (ILs), salts that are liquid close to room temperature, represent another promising class of ionically conducting materials, considered as possible new electrolytes for energy-storage devices as batteries or supercapacitors [8]. The cations (and even some of the anions) forming these salts often have clearly non-spherical, asymmetric shape and, thus, reorientational dynamics must exist in these materials, too. Indeed, in light-scattering (LS) studies of several ILs [9,10,11] ionic reorientational dynamics was detected and revealed to exhibit glasslike freezing, similar to many canonical molecular liquids [12,13]. Ion reorientations in ILs were also detected by nuclear magnetic resonance (NMR) experiments [14,15,16]. However, it remains to be proven whether ionic conductivity in ILs is governed by viscosity or rather dominated by the orientational dynamics.

Dielectric spectroscopy is ideally suited to investigate reorientational dynamics and, thus, it is the most commonly applied method for molecular glass formers and plastic crystals [3,12,17,18]. Dielectric spectroscopy is sensitive to dipolar fluctuations. As most of the cations in ILs have a dipolar moment, their reorientations should be well detectable by this method. There are numerous dielectric studies of ILs but most of them concentrate on the translational dynamics of the ions, only (e.g., Refs. [9,10,11,19,20,21,22,23,24,25,26,27]). This is done by determining the dc conductivity $\sigma_{dc}$, by fitting the spectra [complex conductivity $\sigma^*(\nu)$ and/or permittivity $\varepsilon^*(\nu)$] assuming specific models for hopping charge-transport, or by evaluating the dielectric modulus, $M^* = 1/\varepsilon^*$ [28]. Numerous quantities characterizing the ion dynamics have been obtained along this line [9,10,11,19,20,21,22,23,24,25,26,27], e.g., the characteristic frequency $\omega_{RBM}$ of the random free-energy barrier hopping model (RBM) [29], the conductivity relaxation time $\tau_\sigma$ derived from $M^*$, or the ion-diffusion coefficient $D$. They were compared to each other, to $\sigma_{dc}$, and to results from other experimental methods as the diffusion coefficient deduced from NMR or the viscosity $\eta$. In this way, these dynamic quantities, all essentially reflecting *translational* motions, were found to be more or less closely correlated. However, evaluations of dielectric spectra of ILs in terms of *reorientational* motions are sparse and, in contrast to canonical dipolar liquids [30], for ILs it is still an open question if and how reorientational motions are related to the ionic charge transport.

___________
[*]Corresponding author:: peter.lunkenheimer@physik.uni-augsburg.de

In dielectric spectra, dipolar reorientation processes should give rise to so-called dielectric relaxation which, in the complex permittivity $\varepsilon^* = \varepsilon' - i\varepsilon''$, leads to steps in the dielectric constant $\varepsilon'(\nu)$ and peaks in the loss $\varepsilon''(\nu)$ [12,17,18]. In various dielectric studies of ILs, indeed signatures of relaxation processes were noted but mostly not ascribed to the reorientational ion motions and not treated in detail. Instead, other effects were considered like translational ion hopping [10,11,21,22,23,24,31] correlated anion-cation motions [19], the "reorganization of the ionic atmosphere" after an ion jump [31] or spurious effects arising from insulating impurities [10]. Notable exceptions are investigations at high frequencies far beyond MHz by the groups of Weingärtner and Buchner, where the occurrence of relaxation features was clearly pointed out and ascribed to reorientational motions of the cations (e.g., Refs. [15,32,33,34]). This interpretation was also nicely confirmed by NMR measurements [15]. Unfortunately, these studies, being restricted to temperatures rather close to room temperature only, provide only limited information about the temperature evolution of the reorientational ion dynamics and its connection to translational ionic charge transport. Are these relaxations also seen at lower frequencies? Does this dynamics exhibit glassy freezing? Does it decouple from the well-known glassy dynamics of the translational degrees of freedom? Are these reorientational processes relevant for ionic charge transport? In the present work we intend to solve these open questions by evaluating broadband dielectric spectra for a variety of ILs in terms of reorientational relaxation processes. Our results indicate that the reorientational motions of the ions play an important and so far underestimated role for the ionic charge transport of ILs.

## II. EXPERIMENTAL DETAILS

The IL samples were purchased from IoLiTec (see Table 1 for a list of all investigated ILs and their abbreviations). The purities of most materials was >99%, except for Emim TCM and Bpyr DCA (>98%). Right before the measurements, all samples were dried for several hours in vacuum or $N_2$-gas at elevated temperatures [35]. For the dielectric measurements at frequencies between about $10^{-1}$ and $10^9$ Hz, the liquids were filled into parallel-plate capacitors using glass-fiber spacers with a typical thickness of 0.1 mm. At frequencies up to several MHz, a frequency-response analyzer (Novocontrol Alpha-analyzer) and an autobalance bridge (Agilent 4980A) were used. Measurements at 1 MHz $\leq \nu \leq$ 3 GHz were performed by impedance analyzers (Agilent E4991 or Hewlett-Packard HP4291) using an I-V coaxial technique. Here the sample capacitor is mounted at the end of a coaxial line, bridging inner and outer conductor [12,36]. For cooling and heating, a $N_2$-gas cryostat was employed.

For some ILs, additional differential scanning calorimetry (DSC) measurements were performed using a DSC 8500 from Perkin Elmer. Prior to the DSC measurements, all IL samples were dried for at least 24 h at 373 K in vacuum atmosphere. To determine cooling-rate dependent changes in the fictive temperature (the latter being essentially identical to the rate-dependent $T_g$), we used the method described in Refs. [37,38]. For this purpose, the ILs were quenched with various cooling rates between 0.5 and 100 K/min and, subsequently, reheated with a standard rate of 10 K/min, monitoring their DSC signal. For further details, see Appendix A.

TABLE 1. Ionic liquids investigated in the present work and their abbreviations used in the text and figures.

| Ionic liquid | Abbreviation |
|---|---|
| 1-Methyl-3-octylimidazolium hexafluorophosphate | Omim $PF_6$ |
| 1-Hexyl-3-methylimidazolium hexafluorophosphate | Hmim $PF_6$ |
| 1-Butyl-3-methylimidazolium hexafluorophosphate | Bmim $PF_6$ |
| 1-Butyl-3-methylimidazolium bis-(trifluoromethylsulfonyl)imide | Bmim TFSI |
| 1-Butyl-3-methylimidazolium tetrafluoroborate | Bmim $BF_4$ |
| 1-Butyl-3-methylimidazolium chloride | Bmim Cl |
| 1-Ethyl-3-methyl-imidazolium tricyanomethanide | Emim TCM |
| 1-Butylpyridinium tetrafluoroborate | Bpy $BF_4$ |
| 1-Butyl-1-methylpyrrolidinium dicyanamide | Bpyr DCA |
| (1-Butylpyridinium)$_{0.6}$ (1-Butyl-3-methylimidazolium)$_{0.4}$ tetrafluoroborat | Bpy$_{0.6}$ Bmim$_{0.4}$ $BF_4$ |
| (1-Ethyl-3-methylimidazolium tricyanomethanide)$_{0.33}$ (1-Butyl-3-methylimidazolium chloride)$_{0.67}$ tetrafluoroborate | (Emim TCM)$_{0.33}$ (Bmim Cl)$_{0.67}$ |
| (1-Ethyl-3-methyl-imidazolium)$_{0.33}$ (1-Butyl-3-methylimidazolium)$_{0.67}$ bis-(trifluoromethylsulfonyl)imide | Emim$_{0.33}$ Bmim$_{0.67}$ TFSI |

## III. RESULTS AND DISCUSSION

### A. Dielectric spectra

Figure 1 shows spectra of $\varepsilon'$, $\varepsilon''$, and $\sigma'$ of Bmim $PF_6$ for various temperatures. The covered temperature range extends from the low-viscosity liquid above room temperature deep into the high-viscosity supercooled state close to $T_g \approx 189$ K [27]. Bmim $PF_6$ represents a typical example of an IL with dipolar cation. Qualitatively similar spectra were obtained for another 11 ILs, also including mixtures of different systems [39] and compounds with dipolar anions. As an example for the latter, Fig. 2 shows dielectric spectra of Bmim TFSI.

As commonly found for ionic conductors, at low frequencies and high temperatures the spectra are dominated by electrode-polarization effects [40]. They give rise to huge values of $\varepsilon'$ at low frequencies [Figs. 1(a) and 2(a)] and to a decrease of $\sigma'(\nu)$ at low frequencies [Figs. 1(c) and 2(c)], falling below the level of the intrinsic dc conductivity, signified by the plateau in $\sigma'(\nu)$. Correspondingly, in $\varepsilon''(\nu)$, which is proportional to $\sigma'/\nu$, deviations from the dc-related $1/\nu$ behavior show up at low frequencies [Figs. 1(b) and 2(b)]. These electrode effects arise from the simple fact that, at low frequencies and high temperatures, the ions arrive at the metallic electrodes and form so-called space-charge regions which leads to non-intrinsic Maxwell-Wagner relaxations [40]. In $\sigma'(\nu)$, in addition a succession of a dc plateau, followed by an approximate power-law increase at higher frequencies is observed. Such behavior is often regarded as typical for hopping conductivity in disordered matter and described by various models of electronic and ionic charge transport (e.g., Refs. [29,41,42,43,44]). Indeed, in several works on ILs, dielectric spectra were fitted by a specific hopping model, the RBM (e.g., Refs. [10,22,24,26]).

tions of ILs. The corresponding peak in $\varepsilon''(\nu)$ is obviously superimposed by the strong dc contribution [Figs. 1(b) and 2(b)].

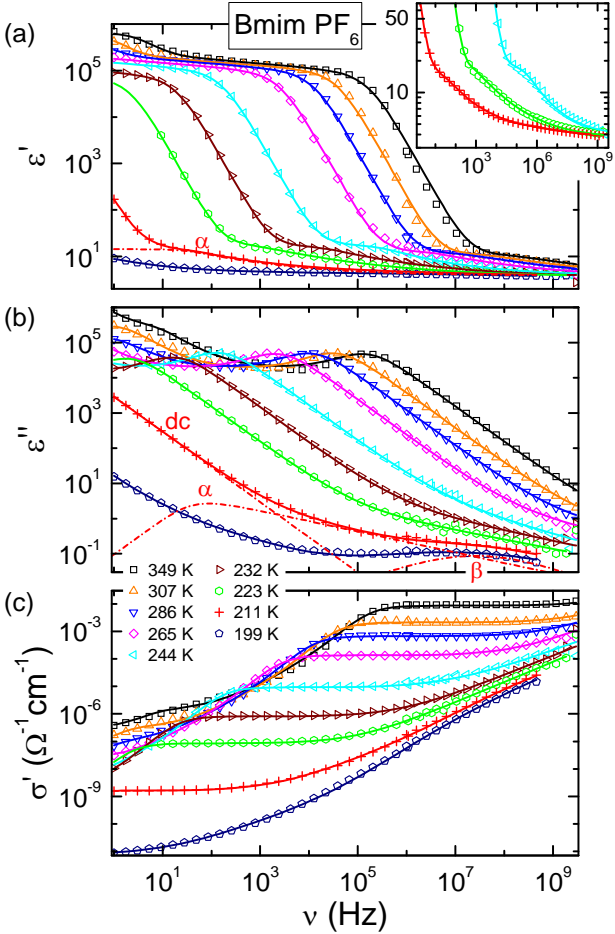

FIG. 1. Frequency dependence of $\varepsilon'$ (a), $\varepsilon''$ (b) and $\sigma'$ (c) of Bmim PF$_6$ at various temperatures. The solid lines are fits as described in the text. $\varepsilon'(\nu)$ and $\varepsilon''(\nu)$ were simultaneously fitted and the $\sigma'$ fit curve calculated from $\varepsilon''$. For 211 K, the dash-dotted lines show the dc, $\alpha$-, and $\beta$-relaxation components of the fit curves. The inset shows a magnified view of $\varepsilon'(\nu)$ in the $\alpha$-relaxation regime.

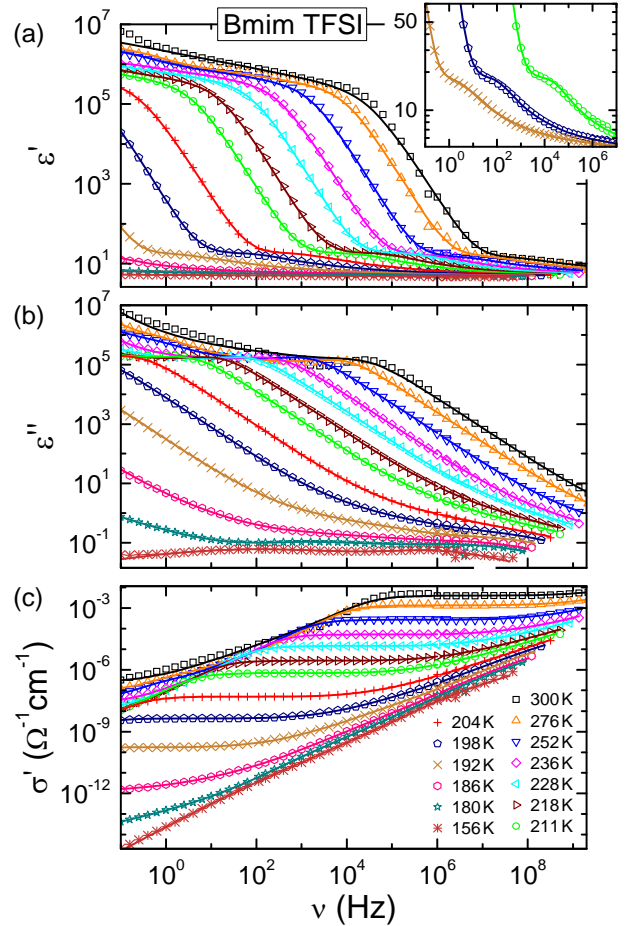

FIG. 2. Frequency dependence of $\varepsilon'$ (a), $\varepsilon''$ (b) and $\sigma'$ (c) of Bmim TFSI, measured at various temperatures. The solid lines are fits as described in the text. The inset shows a magnified view of $\varepsilon'(\nu)$ in the $\alpha$-relaxation regime for three temperatures.

Notably, in nearly all ILs a close inspection of the $\varepsilon'$ spectra reveals a steplike increase with decreasing frequency before the strong increase due to electrode polarization sets in at lower frequencies [10,11,19,21,22,23,24,27,31,39] [cf. dash-dotted line in Fig 1(a) and insets of Figs. 1 and 2]. This reminds of the typical signature of a dipolar relaxation process but it was not treated in detail in most previous investiga-

### B. Interpretation in terms of ion hopping

As mentioned above, in various previous works the RBM was used to fit dielectric spectra on ILs. However, as revealed, e.g., by a careful check of the fit curves in Refs. [10,22,24,26,31], the observed relaxation feature can not or only poorly be described by the RBM, which considers purely translational ionic motions and no dipolar reorientations [29]. As an example, the dashed lines in Fig. 3 and its inset show such fit curves (including an electrode-polarization contribution; see below) for Bmim PF$_6$ at 223 K: While the RBM indeed leads to a relaxation-like step in $\varepsilon'(T)$, which may be ascribed to local ion hopping, the actual experimental data are not well accounted for in this way. When accepting that the relaxation step in the experimental spectra is due to the reori-entation of the ions, of course the RBM model or any other ion-hopping model cannot be expected to properly fit such data.

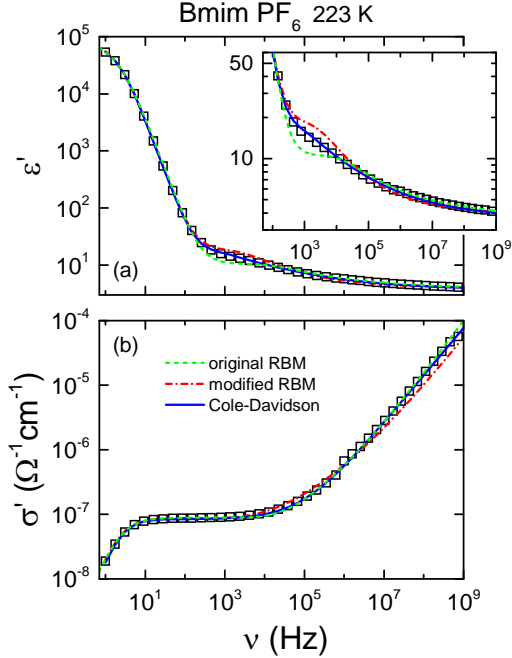

FIG. 3. (a) Frequency dependence of $\varepsilon'$ (a) and $\sigma'$ (c) of Bmim PF$_6$, measured at 223 K. The solid lines are fits assuming a CD function for the description of the reorientational $\alpha$-relaxation as already shown in Fig. 1. The dashed and dash-dotted lines represent alternative fits using the original [29] and modified [45] RBM, respectively. In all cases, the electrode polarization was taken into account by a distributed RC circuit [40]. The inset shows a magnified view of $\varepsilon'(\nu)$ in the $\alpha$-relaxation regime. All fits were simultaneously performed for $\varepsilon'$ and $\sigma'$.

Nevertheless, it cannot be excluded that some of the dozens of other ion-hopping models existing in literature would be able to fit such data. This was explicitly demonstrated for a modified version of the RBM [45], applied to one IL and two polymerized ILs, in a recent work by Gainaru et al. [46]. For the example of Bmim PF$_6$, however, the dash-dotted lines in Fig. 3 demonstrate that only a slightly better fit than with the original RBM can be achieved in this way. Of course, this model only has three parameters and, thus, the fits cannot be expected to be perfect. In any case, one should be aware that the ability or inability to fit such data with a specific ion-hopping model is, in fact, rather meaningless: employing such a model implies a complete neglect of the reorientational dynamics in ILs, although the latter clearly causes the observed relaxation feature in the ILs as discussed in detail in the next section. Overall, it seems to us that the RBM or other ion-hopping models may be suited to describe ionic conductors lacking any permanent dipolar moments, e.g., CKN [47]. However, for ILs the dipolar relaxation from the cations, not accounted for by the RBM, seems to dominate the ac part of the conductivity spectra as discussed in detail below.

In two previous works [10,31], the inability of the original RBM to account for the experimentally observed relaxation feature was also realized and an additional relaxation process was invoked to fit the data. It was ascribed to spurious effects [10] or the "structural reorganization of the ionic atmosphere" [31]. However, one may ask whether it is really necessary to assume any contributions from ionic hopping transport to describe the ubiquitously detected step in $\varepsilon'(\nu)$. Instead, it could merely signify the relaxation process arising from the reorientations of the dipolar ions included in all ILs. To check this idea, in the following, we treat this spectral feature just as the reorientational relaxation processes in canonical dipolar liquids (e.g., glycerol or water) [12,18,48], which can be deconvoluted from the spectra, even for considerable ionic-conductivity contributions [40,49].

### C. Interpretation in terms of reorientational dynamics

To account for the mentioned relaxation feature, we use the empirical Cole-Davidson (CD) function [50], often found to well describe the main reorientational process (termed $\alpha$ relaxation), in dipolar liquids [12,13,18]. In addition, in our experience a proper modeling of the electrode-polarization effects is essential for an accurate evaluation. Here we use an equivalent circuit comprising up to two distributed RC circuits, connected in series to the bulk sample, which is able to fit electrode-polarization effects in ionic conductors with unsurpassed precision [40]. As discussed in detail in Ref. [40], the two RC circuits account for the double steps in $\varepsilon'(\nu)$ as often seen in measurements of ionic conductors extending to sufficiently low frequencies and high temperatures [cf. the 349 K curve in Fig. 1(a)]. Finally, the dielectric spectra of ILs tend to exhibit one or two secondary relaxation processes (sometimes termed $\beta$ relaxation) [24,25,51], a common phenomenon of dipolar liquids [52,53,54]. For example, in Bmim PF$_6$ such a process is nicely revealed by the loss peak observed in Fig. 1(b) around $10^7$ Hz at 199 K and in Bmim TFSI two such processes are observed [at $10^2$ and $10^6$ Hz for 156 K in Fig. 2(b)]. Such secondary relaxations usually can be well approximated by the empirical Cole-Cole (CC) function [55].

Thus, our overall description of the intrinsic sample properties comprises the dc conductivity and, in addition, the sum of a CD and one or two CC functions. In this way, excellent fits of the rather complex broadband spectra shown in Figs. 1 and 2 can be achieved (solid lines), which also is the case for the other 10 investigated ILs listed in Table 1. It should be noted that for none of the spectra at different temperatures, all of the mentioned contributions had to be simultaneously used, which limited the number of fit parameters to reasonable values. For example, at high temperatures the secondary relaxations have completely shifted out of the frequency window and for low temperatures electrode polarization plays no role. Moreover, as seen in Figs. 1 and 2, the different contributions to the spectra are well separated, thus avoiding a correlation of fit parameters. This is also the case for the reorientational relaxation feature that is in the focus of the present work: While its loss peak is superimposed by the dc contribution [cf. the $\alpha$ peak indicated by the dash-dotted line in Fig. 1(c)], in $\varepsilon'(\nu)$ (where dc plays no role) it leads to a significant step with a well-defined point of inflection.

We want to point out that, notably, no frequency-dependent contribution from hopping conductivity was necessary to fit our broadband dielectric data on 12 ILs. Our findings clearly demonstrate that the experimental spectra in ILs are well consistent with a reorientational origin of the observed relaxation features, where the only contribution from ion translation is the dc conductivity. However, admittedly the bare fact that the data can be fitted assuming ion reorientations only can demonstrate that they are consistent with this notion but they cannot finally prove it. Therefore, in the following we compare the obtained relaxation times with results from LS.

reveals a reasonable agreement with the dielectric relaxation times around room temperature reported in Refs. [32,34], which were also ascribed to reorientational motions. A similar comparison of dielectric and LS relaxation times was performed in Ref. [62], also corroborating the reorientational nature of the dielectric relaxation in ILs.

Interestingly, an apparently similar dielectric relaxation feature as in ILs was previously also reported for ionic conductors lacking any permanent dipolar moments [47,63,64]. It was ascribed, e.g., to the "growth and shrinkage of a dipole moment that occurs during the hop" [64] or to the "long range ionic diffusion process" [65] For ILs, Fig. 4 clearly demonstrates a reorientational origin of this spectral feature.

### D. Coupling of the reorientational and translational ion dynamics

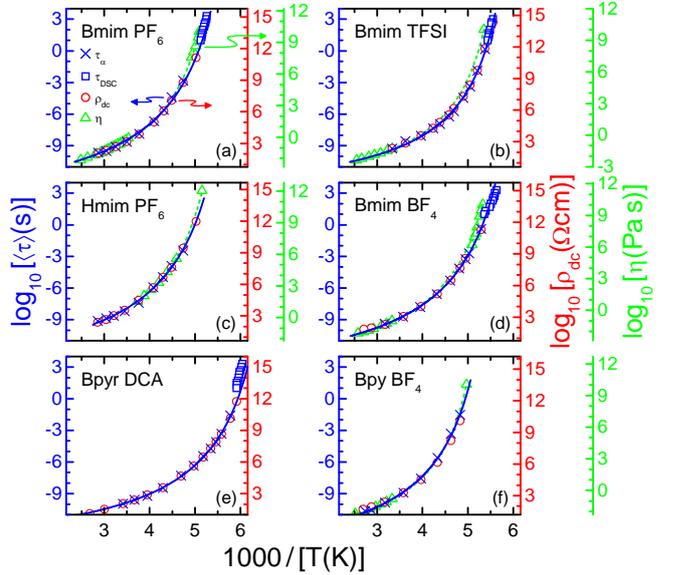

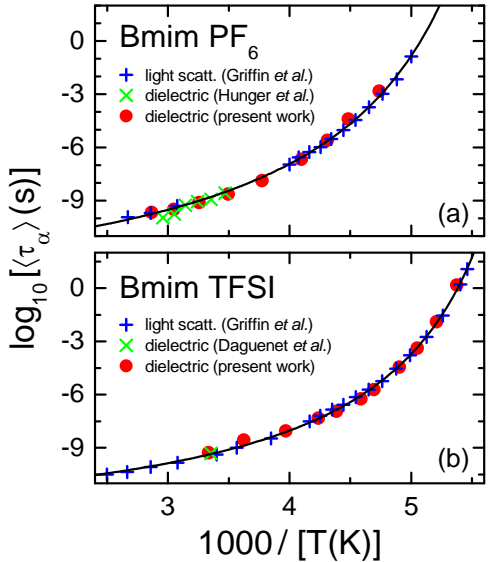

FIG. 4. Temperature-dependent average reorientational $\alpha$-relaxation times of Bmim PF$_6$ (a) and Bmim TFSI (b). Results detected by LS [11] (pluses) and dielectric spectroscopy (circles: present work; crosses: from Refs. [32,34]) are shown. The lines are fits with the VFT formula.

FIG. 5. Comparison of the temperature-dependent translational and reorientational ion dynamics of six ILs. $\rho_{dc}$ (circles; first right scale) and average reorientational relaxation times from dielectric spectroscopy (crosses; left scale) are shown in Arrhenius representation. If available, in addition relaxation times from DSC (squares) and viscosity data [23,66,67,68] (triangles; second right scale) are included. The different ordinates cover the same number of decades. Their starting values were chosen to achieve a good match of the $\rho_{dc}$ and $\langle\tau_\alpha\rangle$ curves (for $\eta$, a match at high temperatures was intended). The lines are VFT fits of $\langle\tau_\alpha\rangle$.

In Fig. 4 for two ILs (Bmim PF$_6$ and Bmim TFSI) the $\alpha$-relaxation times $\tau_\alpha$ determined by us are compared to those deduced from LS [11], which can be unequivocally ascribed to molecular reorientations [11,13,56,57]. (For a brief discussion of the deduced relaxation strength, see Appendix B.) The good agreement indeed proves that the performed analysis and interpretation of the dielectric results in terms of reorientational motions is valid. The lines in Fig. 4 show that $\tau_\alpha(T)$ from both methods can be fitted by the same empirical Vogel-Fulcher-Tammann (VFT) law [58,59,60], as commonly found for supercooled liquids [12,18,61]. Moreover, Fig. 4 also

The results of Fig. 4 nicely confirm the reorientational nature of the observed relaxational response, as also considered in some earlier works [15,27,32,33,34,62]. When accepting this notion, our broadband dielectric data on 12 ILs finally enable far-reaching conclusions on the coupling of the reorientational and translational ion dynamics: The most natural quantity for the characterization of the latter is $\sigma_{dc}$ (or the resistivity $\rho_{dc} = 1/\sigma_{dc}$), well defined by the mentioned plateau in the $\sigma'$ spectra. In contrast to the frequently used parameters $\omega_{RBM}$, $\tau_\sigma$, or $D$, previously deduced from dielectric data [9,10,11,19,20,21,22,23,24,25,26,27,46], this quantity is completely free from any model assumptions. For six ILs, Fig. 5 shows the temperature-dependence of $\rho_{dc}$ (circles) and of $\tau_\alpha$ (crosses) in a common frame using an Arrhenius representation, where the two ordinates were adjusted to cover the same number of decades. By choosing proper starting values of the y-axes, a very good match of the $\rho_{dc}$ and $\tau_\alpha$ curves is achieved. This is also valid for the six other investigated ILs (Fig. 6). This implies a direct proportionality of both quantities as explicitly demonstrated in the double-logarithmic plot of $\rho_{dc}$ vs. $\tau_\alpha$ shown in Fig. 7. There the linear increase with slope one (line in Fig. 7) evidences $\rho_{dc} \propto \tau_\alpha$. Quite unexpectedly, the results for the different investigated ILs roughly follow a common curve with a maximum deviation of about one decade only. This indicates that, rather independent from the partly quite different shapes and sizes of the ions of the investigated ILs, a certain reorientation rate of the ions always induces a similar resistivity value.

dc and ac regimes in the conductivity spectra was considered (which is essentially the same as $\omega_{RBM}$ of the RBM). It was compared to other translational quantities [10,11,21,22,23,24,26,31,62], partly also including the dc conductivity [21,22,23,24]. The frequency $\omega_c$ is sometimes supposed to be identical to the peak frequency in the dielectric loss after subtraction of the dc conductivity, which may lead to the conclusion that it mirrors reorientational motions. However, this notion is solely based on empirical findings in different ionic conductors and there is no principle reason why this should be the case in ILs. Moreover, as pointed out above, within the RBM $\omega_c$ is a purely translational quantity, not related to reorientations. At this frequency, $\sigma'(\nu)$ transfers from the strongly temperature-dependent dc regime to the weakly temperature-dependent ac regime, commonly observed in conductivity spectra of ionic conductors. As demonstrated, e.g., by the $\varepsilon''$ and $\sigma'$ spectra of Figs. 1 and 2, for ILs the ac regime is dominated by the broad relaxation peaks in the loss, which leads to an approximately linear frequency-dependent increase of $\sigma'(\nu)$ [Figs. 1(c) and 2(c)]. Therefore, of course $\omega_c$ should be approximately proportional to the dc conductivity, as indeed was observed [21,22,23,24]. Thus, based on $\omega_c$ it is not possible to draw any conclusion on the connection of reorientational motions and dc charge transport. Only a proper evaluation of the spectra under the assumption of such reorientational motions (justified by the comparison with light scattering; Fig. 4), can reveal unequivocal information on reorientational relaxation times. This implies an evaluation with one of the common relaxation functions as used, e.g., for glycerol or any other dipolar liquid as done in the present work.

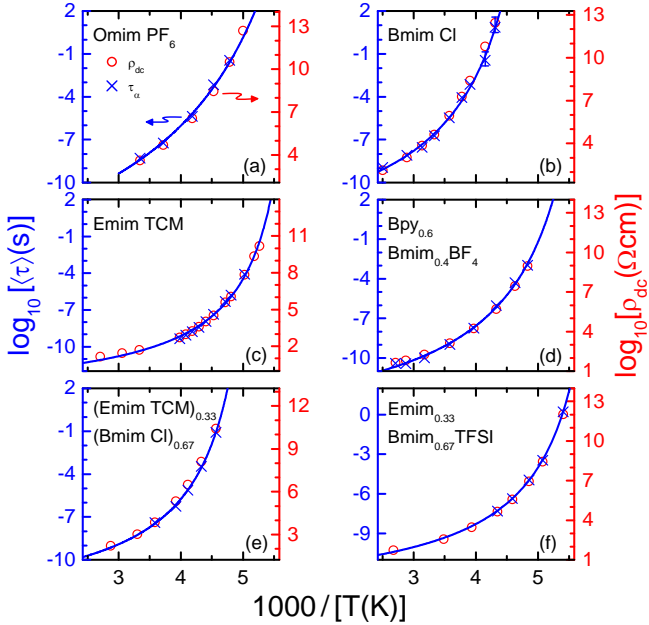

FIG. 6. Same plot as Fig. 5 for the other six investigated ILs.

All these findings clearly demonstrate that the translational and reorientational dynamics of the ions in ILs are very closely coupled. This seems to be a rather general behavior, found in ILs where only the cation is dipolar, where both ions are dipolar, and in IL mixtures. Previous dielectric works on ILs could not arrive at such a conclusion because the data were only evaluated in terms of translational dynamics [10,11,20,21,22,23,24,25,26,27,31,46] or because the investigated temperature range was too restricted to allow for a meaningful comparison of the reorientational and translational ion dynamics [15,32,33,34]. It should be noted that in various earlier works the transition frequency $\omega_c$ between the

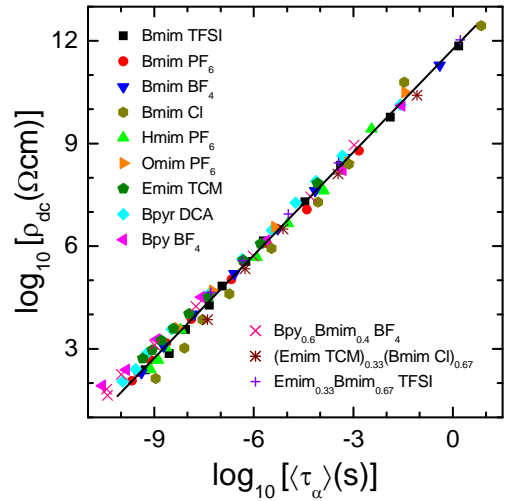

FIG. 7. Dc resistivity vs. the average reorientational relaxation time of various ILs as determined from fits of the dielectric spectra. The line with slope one indicates approximately linear behavior, $\rho_{dc} \propto \langle \tau_\alpha \rangle$.

Remarkably, in Ref. [46] reorientational motions were also considered as possible origin of the observed relaxation feature (however, assuming ion-pair motions instead of reorientations of dipolar ions). Nevertheless, the dielectric spectra in that work were fitted by the RBM, only accounting for translational motions, and, unfortunately, no comparison of reorientational relaxation times with the ionic conductivity was provided. The latter statement can also be made for Ref. [62] where the relaxation was ascribed to a "reorientational step of ions escaping their cage formed by surrounding counterions". The orientational relaxation times deduced by LS, as reported in Ref. [9] for Bmim PF$_6$, were found to be consistent with $\tau_\sigma$ derived from $M''(\nu)$, which is known to be essentially proportional to $\rho_{dc}$, thus corroborating the present results. However, the missing temperature overlap of the $\tau_\alpha(T)$ and $\tau_\sigma(T)$ data in this work limit the significance of this finding. In Ref. [11], a decoupling of $\tau_\alpha(T)$, determined by LS, and the inverse diffusion coefficient was reported for three ILs, including Bmim PF$_6$. However, $D$ was deduced from the RBM, not applicable to dipolar ILs within the present framework, and several additional assumptions had to be made to arrive at absolute values of the diffusion coefficient.

The question arises what is the mechanism of the coupling of the ion rotation dynamics to charge transport found in the present work. Is it of indirect nature, resulting from a close coupling of both dynamics to the viscosity of the material? Or is it a direct coupling, similar to the ionically conducting plastic crystals, where "revolving-door" or "paddle-wheel" mechanisms are considered [6,7]? To help clarifying this issue, in Fig. 5 viscosity data from literature are included for several ILs [23,66,67,68]. They were scaled to match the dielectric results at high temperatures, where decoupling should be least pronounced. Then the $\eta(1/T)$ data seem to exceed the $\rho_{dc}(1/T)$ and $\tau_\alpha(1/T)$ curves at low temperatures but, admittedly, the significance of these rather slight deviations is limited. Interestingly, in Refs. [9] and [20] a similar decoupling of translational ionic motion from viscosity was reported. This implies that, at least at low temperatures, ion reorientation plays a significant role for the ionic charge transport, i.e., at high viscosity, the rotation of the dipolar ions opens paths for the passage of other ions, thus enhancing their mobility.

For several ILs, Fig. 5 also provides relaxation times close to $T_g$, determined by our DSC measurements (see Appendix A for details). Interestingly, except for Bpyr DCA, these data seem to quite closely match the extrapolated $\rho_{dc}(1/T)$ and $\tau_\alpha(1/T)$ curves. This is in accord with the identical temperature characteristics of the dynamics deduced from DSC and dielectric experiments reported in Ref. [37] for several glass-formers. It is obvious that the DSC trace is governed by the thermodynamically dominant processes in a material, which is not necessarily identical with the dynamics determining the viscosity.

## IV. SUMMARY

In the present work, we confirm the interpretation of broadband dielectric spectra of ILs in terms of the often neglected reorientational dynamics of dipolar ions, which leads to the ubiquitous relaxational feature detected in the experimental data. We show that the spectra in 12 ILs can be well described in this way, without invoking any contribution from translational ion transport, apart from the dc conductivity. Our results reveal a close coupling of the ionic charge transport with reorientational motions. Moreover, we find indications for a direct coupling of both dynamics, most likely via a revolving-door mechanism where the rotation of the dipolar ions opens paths for the passage of other ions. A similar mechanism was previously considered for plastic crystals, but, in contrast to these materials, in ILs the centers of the revolving doors can also move translationally. Overall, ion reorientations seem to play a significant, but so far mostly overlooked role for the ionic charge transport in ILs.

## ACKNOWLEDGMENTS

We thank M. Aumüller, A. Fälschle, E. Thoms, and M. Weiß for performing part of the dielectric measurements. This work was supported by the BMBF via ENREKON 03EK3015 and by the Deutsche Forschungsgemeinschaft (Grant No. LU 656/3-1).

## APPENDIX A: DIFFERENTIAL SCANNING CALORIMETRY

Differential scanning calorimetry measures the relative change of a sample's heat capacity via the heat flow (HF) needed to increase/decrease its temperature in comparison to a reference. As mentioned in the main text, we have performed cooling-rate dependent DSC for various ILs to obtain relaxation times in vicinity of the glass transition. Due to its kinetic nature, the actual glass-transition temperature $T_g'$ varies with the applied cooling rate $q$. (The glass temperature $T_g$, as commonly used as a material property, is defined for a fixed cooling rate of 10 K/min.) This transition shows up as a sigmoidally shaped anomaly in the HF under cooling or heating, with additional under- or overshoots, depending on the used cooling and heating rates.

The determination of $T_g'$ from DSC cooling scans comes with a number of disadvantages and can be a problem especially for very fast rates. As explained, e.g., in Refs. [37,38], instead $T_g'$ can be obtained from heating scans with a moderate fixed heating rate, performed after cooling the sample with different rates. From the enthalpy (calculated from the DSC results) the so-called fictive temperature $T_f$, which is essentially identical to $T_g'$, can be deduced [38]. To precisely determine cooling-rate dependent changes in the fictive temperature, we used the method described in Refs. [37,38]. The ILs were quenched with various cooling rates $0.5 < q < 100$ K/min and, subsequently, reheated with the standard rate $q_s$ = 10 K/min. The DSC measurements were performed with a power compensating DSC 8500 from Perkin Elmer. The liquid samples were placed in sealed aluminum pans and for reference empty aluminum pans were used. All ionic-liquid samples were dried for at least 24 h at 373 K in vacuum atmosphere prior to the DSC measurements.

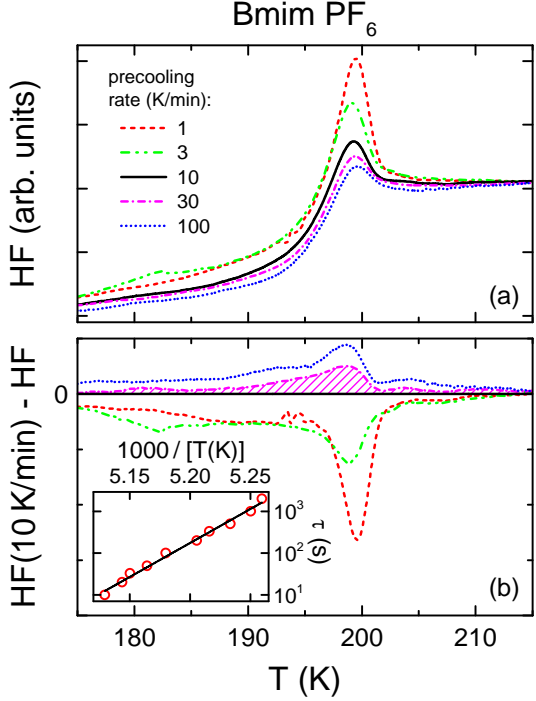

FIG. 8. (a) Temperature dependence of the normalized heat flow (HF) of five 10 K/min heating scans covering the glass-transition region of BMIM PF$_6$. Each heating scan was performed after a different precooling rate ranging from 1 to 100 K/min. (b) Difference in HF to the standard heating scan with 10 K/min. The inset shows the temperature-dependent relaxation times deduced from these DSC data using an Arrhenius representation.

As an example, Fig. 8(a) shows the measured HF for Bmim PF$_6$. The shape of the glass transition step, showing up in the HF during heating, obviously depends on the previously applied precooling rate $q$. Relatively small rates result in an enlarged endothermic contribution sitting on top of the transition step [e.g., red dashed line in Fig. 8(a)] and fast precooling rates produce a reduction of the HF (for very fast rates an exothermic contribution, i.e., a minimum would arise) immediately preceding the transition step [e.g., blue dotted line Fig. 8(a)]. From the heating standard scan, where cooling and heating are performed with $q_s$ [black solid line Fig. 8(a)], a standard fictive temperature $T_f^s$ is determined. Differences of the other heating scans to this standard scan are used to calculate the precooling-rate dependent enthalpy release $\Delta H(q)$ [e.g., hatched area in Fig. 8(b)]. The fictive temperatures $T_f(q)$ corresponding to every precooling rate are then calculated via

$$T_f(q) = T_f^s + \frac{\Delta H(q)}{\Delta C_p}, \qquad (2)$$

where $\Delta C_p$ is the relative change in heat capacity between the glassy and the liquid state [37,38].

Generally, determining the glass temperature by cooling a sample with 10 K/min or by using the condition $\tau(T_g) = 100$ s leads to comparable values. This enables to assign relaxation-time values to the different precooling rates, e.g., 1 K/min corresponds to $10^3$ s and 100 K/min to 1 s. With this, from the rate-dependent fictive temperatures determined by DSC, temperature-dependent relaxation times can be deduced as shown in the inset of Fig. 8(b). The squares in Fig. 5 were also calculated in this way.

## APPENDIX B: RELAXATION STRENGTH

Figure 9 shows the temperature dependence of the relaxation strength of the $\alpha$ relaxation, determined by the fits of the dielectric spectra of Bmim TFSI and Bmim PF$_6$ (see Figs. 1 and 2 and main text). The observed scatter of the data points arises from the fact that the relaxation feature in the spectra is partly superimposed by the contributions from electrode polarization and dc conductivity. Nevertheless, a general Curie like increase of $\Delta\varepsilon(T)$ with decreasing temperature can be clearly identified. In contrast to PF$_6$, the TFSI ion has a dipolar moment. Nevertheless, for both ILs, $\Delta\varepsilon$ is of the same order of magnitude. This reflects the fact that the dominant conformation of TFSI is estimated to have a much lower dipolar moment than the BMIM cation, which is present in both materials [69].

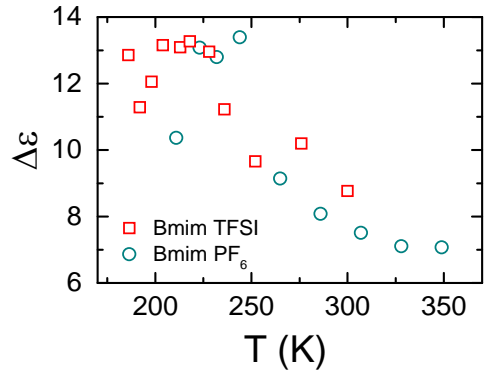

FIG. 9. Temperature dependence of the relaxation strength of the dipolar relaxation process of two ILs as determined from fits of the spectra as described in the main text.

___